\documentclass[12pt,epsf]{article}
\usepackage{graphicx}
\begin{document}

\def\a{\alpha}
\def\b{\beta}
\def\c{\varepsilon}
\def\d{\delta}
\def\e{\epsilon}
\def\f{\phi}
\def\g{\gamma}
\def\h{\theta}
\def\k{\kappa}
\def\l{\lambda}
\def\m{\mu}
\def\n{\nu}
\def\p{\psi}
\def\q{\partial}
\def\r{\rho}
\def\s{\sigma}
\def\t{\tau}
\def\u{\upsilon}
\def\v{\varphi}
\def\w{\omega}
\def\x{\xi}
\def\y{\eta}
\def\z{\zeta}
\def\D{\Delta}
\def\G{\Gamma}
\def\H{\Theta}
\def\L{\Lambda}
\def\F{\Phi}
\def\P{\Psi}
\def\S{\Sigma}

\def\o{\over}
\newcommand{\gsim}{ \mathop{}_{\textstyle \sim}^{\textstyle >} }
\newcommand{\lsim}{ \mathop{}_{\textstyle \sim}^{\textstyle <} }
\newcommand{\vev}[1]{ \left\langle {#1} \right\rangle }
\newcommand{\bra}[1]{ \langle {#1} | }
\newcommand{\ket}[1]{ | {#1} \rangle }
\newcommand{\EV}{ {\rm eV} }
\newcommand{\KEV}{ {\rm keV} }
\newcommand{\MEV}{ {\rm MeV} }
\newcommand{\GEV}{ {\rm GeV} }
\newcommand{\TEV}{ {\rm TeV} }
\def\diag{\mathop{\rm diag}\nolimits}
\def\Spin{\mathop{\rm Spin}}
\def\SO{\mathop{\rm SO}}
\def\O{\mathop{\rm O}}
\def\SU{\mathop{\rm SU}}
\def\U{\mathop{\rm U}}
\def\Sp{\mathop{\rm Sp}}
\def\SL{\mathop{\rm SL}}
\def\tr{\mathop{\rm tr}}

\def\IJMP{Int.~J.~Mod.~Phys. }
\def\MPL{Mod.~Phys.~Lett. }
\def\NP{Nucl.~Phys. }
\def\PL{Phys.~Lett. }
\def\PR{Phys.~Rev. }
\def\PRL{Phys.~Rev.~Lett. }
\def\PTP{Prog.~Theor.~Phys. }
\def\ZP{Z.~Phys. }

\newcommand{\ds}{\displaystyle}
\newcommand{\bear}{\begin{array}}  \newcommand{\eear}{\end{array}}
\newcommand{\bea}{\begin{eqnarray}}  \newcommand{\eea}{\end{eqnarray}}
\newcommand{\beq}{\begin{equation}}  \newcommand{\eeq}{\end{equation}}
\newcommand{\bef}{\begin{figure}}  \newcommand{\eef}{\end{figure}}
\newcommand{\bec}{\begin{center}}  \newcommand{\eec}{\end{center}}
\newcommand{\non}{\nonumber}  \newcommand{\eqn}[1]{\beq {#1}\eeq}
\newcommand{\la}{\left\langle} \newcommand{\ra}{\right\rangle}
\def\lrf#1#2{ \left(\frac{#1}{#2}\right)}
\def\lrfp#1#2#3{ \left(\frac{#1}{#2}\right)^{#3}}


\baselineskip 0.7cm

\begin{titlepage}

\begin{flushright}
IPMU 08-0011
\end{flushright}

\vskip 1.35cm
\begin{center}
{\large \bf
   Gravitational Waves as a Probe of the Gravitino Mass
}
\vskip 1.2cm
Fuminobu Takahashi$^{1}$, T. T. Yanagida$^{1,2}$, and Kazuya Yonekura$^{2}$
\vskip 0.4cm
{\it
${}^1$ Institute for the Physics and Mathematics of the Universe,
  University of Tokyo, Chiba 277-8568, Japan\\
${}^2$ Department of Physics, University of Tokyo, Tokyo 113-0033, Japan
}

\vskip 1.5cm

\abstract{  If gaugino condensations occur in the early universe,
domain walls are produced as a result of the spontaneous breaking of a
discrete $R$ symmetry.  Those domain walls eventually annihilate with
one another, producing the gravitational waves.  We show that the
gravitational waves can be a probe for measuring the gravitino mass,
if the constant term in the superpotential is the relevant source of
the discrete $R$ symmetry breaking.  }
\end{center}
\end{titlepage}

\setcounter{page}{2}

\section{Introduction}
\label{sec:1}

The presence of numerous gauge symmetries is quite a mediocre
phenomenon in the string landscape~\cite{Susskind:2003kw} and it is even
unlikely that our universe possesses only the standard-model (SM)
gauge interactions. Thus, we postulate, through this paper, that we
have many (hidden) gauge bosons besides those in the SM.

Each gauge boson is always accompanied by its fermionic partner called
a gaugino in supersymmetry (SUSY).  Some of the gauginos may condense
in the vacuum by the strong forces of the corresponding gauge
interactions.  Indeed, the gaugino condensation is known to occur in
e.g. a $SU(N)$ pure SUSY Yang-Mills theory.  In the theory, a
continuous $R$ symmetry is explicitly broken to a discrete $Z_{2N}$
subgroup by instantons~\cite{Witten:1982df}.  The gaugino condensation
spontaneously breaks $Z_{2N}$ further down to $Z_{2}$.  Accordingly,
domain walls are formed at epochs of the condensations in the early
universe~\cite{Zeldovich:1974uw,Kibble:1976sj,Vilenkin:1981zs,Dvali:1996xe,
Matsuda:1998ms}.  If the gaugino
condensation occurs before the inflation, the domain walls are diluted
away by the subsequent exponential expansion, and there is no
cosmological remnant of the walls in the observable universe.
However, if the gaugino condensation occurs after the inflation, the
domain walls may play an important role in cosmology, since their
energy density decreases much more slowly than ordinary matters.  In
fact, if the domain walls are stable, they will dominate the energy
density of the universe in the end, which makes the universe extremely
inhomogeneous. To avoid such cosmological disaster, the domain walls
must eventually disappear. Fortunately, we know that the discrete
$Z_{2N}$ symmetry is not an exact symmetry and hence the domain walls
are not stable.

In this paper, we discuss the dynamics of the domain walls formed by
the gaugino condensation, assuming that the $SU(N)$ gauge sector 
does not have sizable couplings to the other sectors. We will briefly discuss
the case that such couplings to the SUSY breaking sector are present later.
Then the source of the breaking of the $R$ symmetry relevant for the
$SU(N)$ sector comes only from the constant term $w_0$
in the superpotential $W$.  The constant term preserves only the $Z_2$
symmetry, and it is needed to cancel the cosmological constant in
supergravity (SUGRA). We find that the walls collapse at $H\simeq
m_{3/2}$, where $H$ is the expansion rate of the universe and
$m_{3/2}$ the gravitino mass.  Interestingly, gravitational
waves~\cite{GW} are likely produced during the violent annihilation
processes of the domain walls~\cite{Gleiser:1998na}. The typical
frequency of the gravitational waves we observe today depends on the
gravitino mass, while the intensity is sensitive to the gaugino
condensation scale, $\Lambda$. Moreover, thermal history after
inflation affects both the frequency and the intensity.  We point out
in this paper that such stochastic gravitational waves may be
detectable by the forthcoming experiments such as Advanced
LIGO~\cite{Barish:1999vh,Fritschel:2003qw}, LCGT~\cite{Kuroda:2002bg},
LISA~\cite{LISA}, BBO~\cite{Crowder:2005nr} and
DECIGO~\cite{Seto:2001qf}, for wide ranges of the gravitino mass and
the reheating temperature.

\section{Gaugino condensation and domain walls}
\label{sec:2}
We briefly explain the gaugino condensation and the subsequently
formed domain walls. Let us consider a pure $SU(N)$ SUSY Yang-Mills
theory~\footnote{One may end up with a pure SUSY Yang-Mills theory at
a low-energy scale, after integrating out heavy flavors.}.  Then the
gauge interactions become strong at a dynamical scale $\Lambda$, and
the gaugino condensation occurs:
\beq
\label{gc}
\la\lambda^a \lambda^a \ra_k \;=\; -32 \pi^2 e^{2 \pi i k/N} \Lambda^3,
\eeq
where $k = 1, \dots,N$, and there are $N$ degenerate vacua.  The
effective superpotential below the dynamical scale is given by
\beq
W_{GC} \;=\; N \Lambda^3 e^{2 \pi i k/N}.
\eeq
One can derive the forms of the above expressions using the symmetry
and the holomorphy of the superpotential.

We assume that the universe undergoes an exponential expansion called
inflation. After inflation ends, the inflaton continues to dominate
the energy density of the universe until it reheats the universe by
the decay.  Even before the reheating, there is a background thermal
plasma.  The maximal temperature of the background plasma can be
expressed in terms of the reheating temperature $T_R$ and the Hubble
parameter during inflation $H_{\rm inf}$ as~\cite{Asaka:1999xd,Giudice:2000ex}
\beq
T_{\rm max} \sim \left(T_R^2 \, H_{\rm inf} M_P\right)^\frac{1}{4},
\eeq
where $M_P = 2.4 \times 10^{18}$ GeV is the reduced Planck mass. 
We require that 
the maximal temperature of the background plasma is higher than the dynamical scale:
\beq
T_{\rm max} \;\gsim\; \Lambda.
\label{eq:tmax}
\eeq
Then, the non-zero gaugino condensate is formed after inflation, when
the cosmic temperature becomes comparable to $\Lambda$~\footnote{Here
and in what follows we assume that the $SU(N)$ gauge sector is in
thermal equilibrium with a temperature comparable to that of the SM
sector.  If the $SU(N)$ gauginos are not in thermal equilibrium,
gaugino condensation occurs when the Hubble parameter, not the
temperature, becomes comparable to $\Lambda$. For the domain walls to
be formed, $H_{\rm inf} \gsim \Lambda$ must be satisfied.  Even in
this case, if the domain-wall network reaches the scaling regime, our
arguments remain essentially unchanged.}.  When gaugino condensation
occurs, one of the $N$ vacua is selected in each domain with the size
of $O(\xi)$, where $\xi$ denotes the correlation length, $\xi\sim
T^{-1}$.  Since the correlation length is much smaller than the
horizon scale at the formation, there appear domain walls that connect
the distinct vacua populated in each domain. The properties of the
domain walls have been extensively studied (see
Refs.~\cite{Dvali:1996xe,Kovner:1997ca}). In particular, the tension
of the wall connecting the adjacent vacua is analytically given
by~\cite{Dvali:1996xe}
\bea
\sigma &=&  2N \Lambda^3  \left| e^{2\pi i/N} - 1\right|.
\label{tension}
\eea
In the limit of large $N$, $\sigma$ approaches $4 \pi \Lambda^3$.  In
the following analysis we adopt $\sigma = 4 \pi \Lambda^3$ as a
reference value.

How does the domain-wall network evolve? The
numerical~\cite{Press:1989yh,Coulson:1995nv,Larsson:1996sp} and
analytical~\cite{Hindmarsh:1996xv} results obtained so far show that
the domain walls reach a scaling regime in simple models. We therefore
assume that the evolution of the domain walls is given by the scaling
law~\footnote{We assume that there is no light particles in plasma
which strongly couple to the domain walls.}.  That is, an averaged
number of the walls per horizon remains the same in the evolution of
the universe.  This should be the case if the annihilation processes
of the domain walls are efficient. The energy density of the domain
walls, $\rho_{DW}$, obeying the scaling law is given by
\beq
\rho_{DW} \;\simeq\;  n\, \sigma \,H,
\label{scaling-law}
\eeq
where $H$ is the Hubble parameter, $n$ denotes an averaged number of
the walls per horizon, and we expect $n \sim O(1)$. Note that the
energy density of the domain walls decreases much more slowly than the
radiation and ordinary matters.  Therefore, if the domain walls are
stable, they come to dominate the energy density of the universe when
\beq
H \;\sim\; \frac{\sigma}{M_P^2}.
\eeq
Once the domain walls dominate the energy density of the universe, the
universe will become intolerably inhomogeneous and
anisotropic~\cite{Vilenkin:1981zs}.  Therefore the domain walls must
disappear before it starts to dominate the universe.

One of the easiest ways to induce the decay of the domain walls is to
introduce a bias and lift the degeneracy of the vacua. That is, if the
spontaneously broken discrete symmetry is only approximate, the false
vacua will decay into the true vacuum in the end, and the domain walls
will disappear.  Such biased domain walls was first studied by
Vilenkin~\cite{Vilenkin:1981zs}.  Here we summarize the result. Let $\epsilon$ denote the
bias between the false and true vacua. ($\epsilon$ has mass dimension
$4$.) Then the domain walls annihilate with one another and disappear
when the tension and the pressure on the wall due to the bias become
comparable.  The Hubble parameter at the decay is
\beq
H_{\rm decay} \;\sim\; \frac{\epsilon}{ \sigma}.
\label{hdecay}
\eeq
Since the decay must be complete before the walls come to dominate the
universe, there is a lower bound on the bias:
\beq
\epsilon \;\gsim\; \frac{\sigma^2}{M_P^2}.
\label{const1}
\eeq
On the other hand, if the bias is too strong, the domain walls are not
formed, or even if formed, the walls might disappear before they reach
the scaling law.  We require that the bias is much smaller than
$\rho_{DW}$ when they reach the scaling law soon after the
formation. That is,
\beq
\sigma H_f \;\gg\; \epsilon
\label{const2}
\eeq
with 
\beq
H_f \;\sim\;\left\{
\bear{cc}
\ds{\frac{\Lambda^2}{M_P}}&{\rm ~~~for~~}T_R > \Lambda \\
\ds{\frac{\Lambda^4}{T_R^2 M_P}}&{\rm ~~~for~~}T_R < \Lambda 
\eear
,\right.
\eeq
where $H_f$ denotes the Hubble parameter at the domain-wall formation.

Our concern here is if such a bias exists in the case of the gaugino
condensation.  Fortunately, there is indeed a source for the explicit
breaking of the discrete $Z_{2N}$ symmetry. In order to cancel the
cosmological constant, we need to introduce a constant $w_0$ in the
superpotential. The constant explicitly breaks $Z_{2N}$ down to $Z_2$,
and so, the degeneracy of the vacua is lifted. To see this, we write
down the scalar potential in SUGRA using the effective superpotential
$W = W_{GC} + w_0$~\footnote{
The non-zero value of $w_0$ produces the gaugino mass,
$m_\lambda = (3N g^2/16\pi^2)  m_{3/2}$,  by the anomaly
mediation~\cite{AMSB}, where $g$ is the gauge coupling of $SU(N)$.
One can also derive the effective potential, (\ref{eff-pote}), from
the gaugino mass term with the
gaugino condensation~\cite{Kovner:1997ca}. 
},
\beq
V \;\simeq\; -3 N  \frac{w_0^*}{M_P^2} \Lambda^3 e^{2 \pi i k/N} + {\rm h.c.}.
\label{eff-pote}
\eeq
Thus, depending on the relative phase between the gaugino condensate
and $w_0$, the scalar potentials take different values in the distinct
$N$ vacua.  The typical size of the bias between the adjacent vacua is
given by
\beq 
\epsilon \;\simeq\; \alpha\, m_{3/2} \Lambda^3, 
\label{bias-gc}
\eeq
where we have used $|w_0| = m_{3/2} M_P^2$, and $\alpha = O(10)$ is a
numerical coefficient. Throughout this paper, we assume that $w_0$ is
the dominant source of the bias.  Substituting (\ref{tension}) and
(\ref{bias-gc}) into (\ref{hdecay}), one can see that the Hubble
parameter at the decay is given by
\beq
H_{\rm decay}\;\sim\; m_{3/2}.
\eeq
Thus, as long as the constant $w_0$ is the main source of the bias,
the timing of the domain-wall decay is determined by the gravitino
mass.  The constraints on the bias, (\ref{const1}) and (\ref{const2}),
read
\bea
\Lambda_{\rm min}^3
 \;\ll\; \Lambda^3 \;\lsim\; 0.1\, m_{3/2} M_P^2 
\label{bound-on-Lambda}
\eea
with 
\beq
\Lambda_{\rm min} \;\equiv\; \left\{
\bear{cc}
\ds{\sqrt{m_{3/2} M_P}}&{\rm ~~~for~~}T_R > \Lambda \\
\ds{\left(m_{3/2}M_P T_R^2\right)^\frac{1}{4}}&{\rm ~~~for~~}T_R < \Lambda 
\eear
\right..
\eeq
As we see in the next section, the intensity of the gravitational
waves increases as $\Lambda$ becomes large. When $\Lambda$ takes the
largest allowed value, the energy of the domain walls becomes
comparable to the total energy of the universe at the decay.

The energy density stored in the domain walls must go into light
degrees of freedom. In this paper we simply assume that the energy
quickly goes into another hidden massless $U(1)$ gauge bosons through
loop diagrams of some heavy vector-like quarks charged under both
gauge symmetries~\footnote{We assume that the effective 
couplings between the $U(1)$ and $SU(N)$ sectors, obtained after the
integration of the heavy quarks,  are so weak that the domain-wall dynamics
is not affected. }. Since we assume that the domain walls decay before
they come to dominate the universe, the energy of the hidden radiation
is subdominant compared to the visible one. In Sec.~\ref{sec:4}, we
will discuss a possibility to observe the effect of the hidden
radiation.  Even if the domain walls decay into the visible particles,
the following arguments on the gravitational waves remain intact, as
long as the decay proceeds fast enough.

Lastly let us summarize our picture.  The domain walls are formed when
the gauginos condense, i.e., $T \sim \Lambda$. Those walls evolve
according to the scaling law (\ref{scaling-law}) until $H \sim
m_{3/2}$.  When $H \sim m_{3/2}$, the domain walls experience the
pressure due to the bias, and they are pulled toward one another.
After violent collisions of the domain walls, the false vacua decay
into the true vacuum, and the domain walls disappear. The energy
stored inside the domain walls and the false vacua will be released
into light degrees of freedom.  Most importantly, since the walls are
spatially extended two dimensional objects, the collisions likely
occur in such a way widely deviated from the spherical symmetry,
leading to the production of the gravitational waves.  Thus the domain
walls of the gaugino condensate may have left their traces in the
gravitational waves, which will be discussed in the next section.

\section{Gravitational waves}
\label{sec:3}
We discuss here the properties of the gravitational waves produced
from the decay of the domain walls.  To detect the gravitational waves
by the forthcoming experiments, the frequency and the intensity must
be in certain ranges.

Let us first consider the frequency of the gravitational waves.  The
gravitational waves are expected to peak at the frequency
corresponding to a typical physical scale when they are produced. In
our case, this frequency corresponds to the horizon scale at the
decay, i.e., $f_* \sim m_{3/2}$. The frequency is red-shifted due to
the subsequent cosmic expansion, and the frequency we observe today,
$f_0$, is much smaller than $f_*$:
\beq
f_0\;=\;\frac{a_*}{a_0}\, f_*,
\eeq
where $a_0$ and $a_*$ are the scale factors at present and at the
formation of the gravitational waves, respectively.  The amount of the
red-shift, or equivalently, $a_*/a_0$, depends on the thermal history
of the universe.

If the reheating of the universe is completed before $H \sim m_{3/2}$,
the universe is radiation-dominated afterwards until the
matter-radiation equality. The ratio of the scale factors is estimated
by
\beq
\frac{a_*}{a_0} \;\simeq\; 6 \times 10^{-14} 
\lrfp{g_*}{200}{-\frac{1}{3}} \lrfp{T_*}{1{\rm\, GeV}}{-1},
\eeq
where $T_*$ denotes the cosmic temperature, and $g_*$ counts the
relativistic degrees of freedom, both evaluated at $a=a_*$.  Thus the
frequency at present, $f_0$, is directly related to the gravitino mass
as
\bea
f_0  &\simeq&0.1\, {\rm Hz}\, \lrfp{g_*}{200}{-\frac{1}{12}} 
          \lrfp{m_{3/2}}{1{\rm\, keV}}{\frac{1}{2}}.
\label{freq1}          
\eea

On the other hand, if the reheating is not completed when $H \sim
m_{3/2}$, the universe is matter-dominated until the reheating. Then
the amount of the red-shift is modified as
\beq
\frac{a_*}{a_0} \;\simeq\; 1 \times 10^{-21} 
\lrfp{T_R}{1{\rm\, GeV}}{\frac{1}{3}} \lrfp{m_{3/2}}{1{\rm\, keV}}{-\frac{2}{3}}.
\eeq
The present frequency becomes
\beq
f_0 \;\simeq\; 2 \,{\rm mHz}\, \lrfp{T_R}{1{\rm\, GeV}}{\frac{1}{3}} 
          \lrfp{m_{3/2}}{1{\rm \,keV}}{\frac{1}{3}}.
\label{freq2}          
\eeq
Surprisingly, the peak frequencies of the gravitational waves coincide
with those covered by various ongoing and planned experiments, for
wide ranges of the gravitino mass and the reheating temperature.  We
will come to this point later.

The intensity of the gravitational waves decreases as the universe
expands, since the amplitude is red-shifted for a mode with the
wavelength less than the horizon size.  In order to characterize the
intensity, one may use a dimensionless quantity $\Omega_{\rm gw}(f)$
defined by
\beq
\Omega_{\rm gw}(f) \;\equiv\; \frac{1}{\rho_c} \frac{d \rho_{gw}}{d \log f},
\eeq
where $\rho_{gw}$ is the energy density of the gravitational waves,
$\rho_c$ the critical energy density, and $f$ the frequency.

Let us estimate the magnitude of $\Omega_{\rm gw}(f)$ (see \cite{GW}).
The energy of the gravitational waves in a horizon at $H = m_{3/2}$ is
estimated by
\beq
E_{gw} \;\sim\; G\, \frac{M_{DW}^2}{R_*} \sim  \frac{2 \pi \Lambda^6}{m_{3/2}^3 M_P^2},
\label{egw}
\eeq
where $G=1/(8 \pi M_P^2)$ is the Newton constant, $M_{DW}$ the energy
stored in the domain walls, and $R_*$ the typical spatial scale of the
energy distribution.  In the second equality in (\ref{egw}), we have
used $M_{DW} \sim \sigma / m_{3/2}^2$ and $R_* \sim 1/m_{3/2}$.  If
the reheating is completed before $H = m_{3/2}$,
we have
\bea
\Omega_{\rm gw}(f_0) h^2 &\sim& 10^{-7}\lrfp{g_*}{200}{-\frac{1}{3}}  
\lrfp{\Lambda^3}{0.1\,m_{3/2} M_P^2}{2},
\label{omega1}
\eea
where $f_0$ is given by (\ref{freq1}), and $h$ is the present Hubble
parameter in units of $100$\,km/s/Mpc. For $\Lambda$ close to its
upper bound given by (\ref{bound-on-Lambda}), therefore, the energy
density of the gravitational wave amounts to one percent of the cosmic
microwave background photons!  If the reheating is not completed at $H
= m_{3/2}$, the gravitational waves are diluted by the entropy
production of the inflaton. The density parameter, $\Omega_{\rm gw}$,
is then given by
\beq
\Omega_{\rm gw}(f_0) h^2 \;\sim\; 10^{-12} \lrfp{T_R}{100{\rm\,GeV}}{\frac{4}{3}} 
								      \lrfp{m_{3/2}}{1{\rm\,keV}}{-\frac{2}{3}}
								       \lrfp{\Lambda^3}{0.1m_{3/2} M_P^2}{2},
\label{omega2}								
\eeq
where $f_0$ is given by (\ref{freq2}). Note that the intensity of the
gravitational waves is very sensitive to the dynamical scale,
$\Lambda$. In order to detect the gravitational waves from the
domain-wall decay, $\Lambda$ must be close to its upper bound.

We have made a rough estimate of the intensity of the gravitational
waves in the above.  Further detailed analyses on the domain-wall
dynamics may reveal the presence of additional suppression
factors. However, there are at least two reasons to believe that the
gravitational waves produced by the domain-wall dynamics are not
extremely suppressed.  One is that the collisions among the domain
walls are strongly first order phase transition, since they are
decoupled from the thermal bath.  The other is that the domain-walls
are spatially extended two dimensional objects, which exhibits a
peculiar dependence on the spatial scale.  To see this more
explicitly, let us consider the dependence of $\Omega_{\rm gw}$ on the
spatial scale, $R$.  Using $M_{DW} \propto R^2$, we have $E_{gw}
\propto R^3$, leading to $\Omega_{\rm gw} \propto R^0$!  This is very
different from the ordinary non-relativistic matters, which would lead
to $\Omega_{\rm gw} \propto R^2$. Therefore, the resultant intensity
of the gravitational waves is independent of the spatial scale, which
makes the above rough estimate more robust~\footnote{The peak
frequency scales as $R^{-1}$, though.}.  Thus, the domain-wall
dynamics may be an ideal source for generating the gravitational waves
in the early universe.

Here let us briefly mention the sensitivities of the ongoing and
planned experiments on the gravitational waves. One of the
ground-based experiments, LIGO~\cite{Abramovici:1992ah}, is in
operation and it is sensitive to the frequency between $O(10)$\,Hz and
$10^4$\,Hz. The latest upper bound is $\Omega_{\rm gw} h^2 < 6.5
\times 10^{-5}$ around $100$\,Hz~\cite{recent-LIGO}, and an upgrade of
the experiment, Advanced LIGO, would reach sensitivities of
$O(10^{-9})$. The sensitivity of LCGT would be more or less similar to
that of Advanced LIGO.  There are also planned space-borne
interferometers such as LISA, BBO and DECIGO. LISA is sensitive to the
band of $(0.03-0.1)\,{\rm mHz} \lsim f_0 \lsim 0.1$\,Hz, and it can
reach $\Omega_{\rm gw} h^2 < 10^{-12}$ at $f_0 = 1$\,mHz.  Moreover, BBO
and DECIGO will cover $10{\rm\,mHz} \lsim f_0 \lsim 10^2$\,Hz with much
better sensitivity.

\begin{figure}[t!]
\begin{center}
\includegraphics[width=12cm]{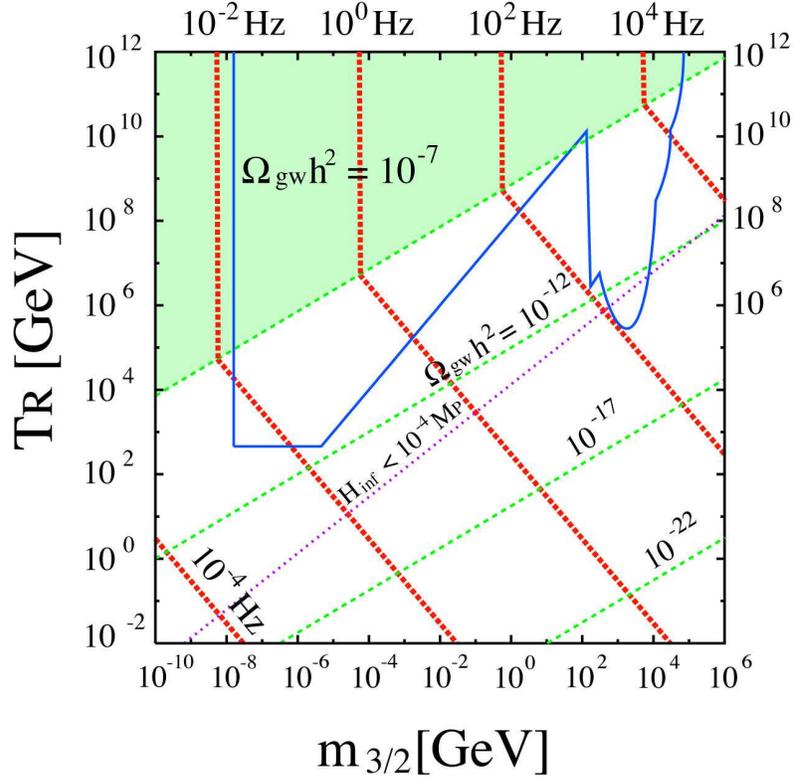}
\caption{
Contours of the frequency (thick-dotted, red) and the maximal intensity (thin-dashed, green)
of the gravitational waves from the domain-wall decay. $\Lambda$ is set to be the largest
 value allowed by (\ref{bound-on-Lambda}).
 The solid (blue) line schematically
shows upper bounds on the reheating temperature from the gravitino problem.
The dotted (purple) line shows the bound coming from (\ref{eq:tmax}) and 
$H_{\rm inf} < 10^{-4}\,M_P$, and the region below the dotted (purple) line is excluded.
Note that the bound is not applied if the $SU(N)$ gauginos are not in thermal equilibrium.
LIGO and LISA are sensitive to the bands of $O(10){\rm\,Hz} \lsim f_0 \lsim 10^4$\,Hz
and $0.1{\rm\,mHz} \lsim f_0 \lsim 0.1$\,Hz, respectively, while BBO and DECIGO
will cover $10{\rm\,mHz} \lsim f_0 \lsim 10^2$\,Hz with much better sensitivity.
}
\label{fig:mgtr}
\end{center}
\end{figure}

In Fig.~\ref{fig:mgtr}, we show the contours of the frequencies: $f_0 =
10^{-4}$, $10^{-2}$, $1$, $10^2$, and $10^4$\,Hz, and the (maximal)
intensities: $\Omega_{\rm gw}h^2 = 10^{-22}, 10^{-17}, 10^{-12}$ and $10^{-7}$,
in the parameter space of the gravitino mass and the reheating
temperature.  Here we have set $\Lambda$ to be the largest value
allowed by (\ref{bound-on-Lambda}).  In the shaded triangle region,
$\Omega_{\rm gw}h^2$ is saturated to $10^{-7}$, since the reheating is
complete before the domain-wall decay (see (\ref{freq1}) and
(\ref{omega1})).  Also we schematically show the upper bounds on $T_R$
from the gravitino problem~\cite{Kawasaki:2004yh, Kohri:2005wn,Jedamzik:2006xz, Moroi:1993mb,Bolz:1998ek,Bolz:2000fu,Ellis:2003dn,Steffen:2006hw,Pradler:2006qh} (solid (blue) line). The dotted (purple)
line shows the constraint obtained by (\ref{eq:tmax}) and $H_{\rm inf}
< 10^{-4}\,M_P$~\cite{Spergel:2006hy}, and the region below the line is
excluded~\footnote{If the $SU(N)$ sector is not thermalized, the
constraint of the dotted (purple) line is not applied.}.

One can see that, for wide ranges of $m_{3/2}$ and $T_R$, the peak
frequencies of the gravitational waves correspond to those covered by
the ongoing and planned experiments.  Compared with the sensitivities
of the experiments mentioned above, the gravitational waves from the
domain-wall decay may be detectable, even by Advanced LIGO.  Once it
is detected, we will be able to obtain information on both the
gravitino mass and the reheating temperature. In particular, for
$m_{3/2} \lsim O(10)$\,eV, the measured frequency can be directly
related to the gravitino mass by (\ref{freq1}).

\section{Discussion and conclusions}
\label{sec:4}
We have so far focused on the gravitational waves produced by the
domain-wall dynamics. Let us here briefly discuss the decay products
of the domain walls. After the collisions of the domain walls, the
energy stored in the walls is assumed to decay into the hidden
massless $U(1)$ gauge bosons. Then there is a relation between the
energies of the hidden radiation and the gravitational waves:
\beq
\Omega_{hid}  \sim 2 \left(\frac{m_{3/2} M_P^2}{\Lambda^3}\right)
				\, \Omega_{\rm gw},
\label{hid-gw}
\eeq
where $\Omega_{hid}$ is the density parameter of the hidden
radiation. Therefore, the hidden radiation is always more abundant
than the gravitational waves (see (\ref{bound-on-Lambda})).  For the
largest allowed value of $\Lambda$, we have $\Omega_{hid} \sim 20\,
\Omega_{\rm gw}$. It might be possible to measure $\Omega_{hid}$
indirectly by using the big bang nucleosynthesis (BBN), the cosmic
microwave background, and large-scale structure data. Indeed,
measurements of the light-element abundances give $\Omega_{hid}h^2
\lsim 1.5 \times 10^{-5}$ at 95\% C.L.~\cite{Mangano:2006ur}, which is
translated to the bound on the gravitational waves,
\beq
\Omega_{\rm gw}h^2 \;\lsim\; 8 \times 10^{-6} \frac{\Lambda^3}{m_{3/2} M_P^2}.
\label{bbn}
\eeq
Therefore, if the domain walls mainly decay into the hidden massless
$U(1)$ gauge bosons, the indirect measurement of $\Omega_{hid}$ 
can set a bound on $\Omega_{\rm gw}$~\footnote{It should be noted however
that the presence of the hidden radiation is not an inevitable
consequence. If the domain walls mainly decays into the visible
particles, the bound (\ref{bbn}) is not applied.}.

If SUSY is realized at a scale much higher than the weak scale,
it may be difficult to directly produce SUSY particles in the accelerator experiments
like LHC or ILC. Then, it will be interesting if we can probe high-energy physics
inaccessible to the accelerator experiments, by observing the gravitational waves.
For $m_{3/2} \sim 10^4 - 10^6$\,GeV, the frequency of the gravitational waves 
is in the band of  $10^2{\rm\,Hz} \lsim f_0 \lsim 10^4$\,Hz, where
the ground-based experiments are sensitive (see Fig.~\ref{fig:mgtr}). If the
gravitino mass is heavier than $10^6$\,GeV, however, $f_0$ exceeds the frequencies 
to which the ongoing and planned experiments are sensitive.
Even in this case, one may be able to indirectly observe the hidden radiation 
related to the gravitational waves by (\ref{hid-gw}).

We have assumed that the constant in the superpotential, $w_0$, is the
relevant source of the discrete $R$ symmetry breaking. 
If the $SU(N)$ gauge sector has couplings to the SUSY breaking sector,
the gaugino mass $m_{\lambda}$ generically receives contribution larger 
than the anomaly mediation.  In that case the relation between $f_0$ and $m_{3/2}$ will 
be lost. However, the properties of the gravitational waves we have derived in the previous section
are still valid if one replaces the gravitino mass by the $(16\pi^2/3N g^2) m_\lambda$.

Gaugino condensation is ubiquitous, if there are many gauge groups
besides the SM ones in nature. The domain walls are generically formed
when the gauginos condense. To avoid cosmological disaster, however, 
those domain walls must disappear.   Since the discrete $R$ symmetry is explicitly
broken by the constant term $w_0$ in the superpotential,  the domain
walls are unstable and eventually decay into light particles. 
Interestingly, the gravitational waves are generated at the
domain-wall collisions, and they contain valuable information on high-energy physics
and the early universe.  In this paper we have pointed out that the resultant
gravitational waves can be a probe for measuring the gravitino mass,
if $w_0$ is the relevant source of the discrete $R$ symmetry breaking.

\section*{Acknowledgment}
F.T. would like to thank K. Nakayama for useful discussions.
This work was supported by World Premier International Research Center
Initiative (WPI Program), MEXT, Japan.


\begin{thebibliography}{99}

\bibitem{Susskind:2003kw}
  L.~Susskind,
  arXiv:hep-th/0302219.
  
\bibitem{Witten:1982df}
  E.~Witten,
  Nucl.\ Phys.\  B {\bf 202}, 253 (1982).
  
\bibitem{Zeldovich:1974uw}
  Y.~B.~Zeldovich, I.~Y.~Kobzarev and L.~B.~Okun,
  Zh.\ Eksp.\ Teor.\ Fiz.\  {\bf 67}, 3 (1974)
  [Sov.\ Phys.\ JETP {\bf 40}, 1 (1974)].
 
\bibitem{Kibble:1976sj}
  T.~W.~B.~Kibble,
  J.\ Phys.\ A  {\bf 9}, 1387 (1976).
  
\bibitem{Vilenkin:1981zs}
  A.~Vilenkin,
  Phys.\ Rev.\  D {\bf 23} (1981) 852.

\bibitem{Dvali:1996xe}
  G.~R.~Dvali and M.~A.~Shifman,
  Phys.\ Lett.\  B {\bf 396}, 64 (1997)
  [Erratum-ibid.\  B {\bf 407}, 452 (1997)].
  
\bibitem{Matsuda:1998ms}
  T.~Matsuda,
  Phys.\ Lett.\  B {\bf 436}, 264 (1998).

\bibitem{GW}
See 
  M.~Maggiore,
  Phys.\ Rept.\  {\bf 331}, 283 (2000), for a review.

\bibitem{Gleiser:1998na}
  M.~Gleiser and R.~Roberts,
  Phys.\ Rev.\ Lett.\  {\bf 81}, 5497 (1998).
  
  
 
\bibitem{Barish:1999vh}
  B.~C.~Barish and R.~Weiss,
  Phys.\ Today {\bf 52N10} (1999) 44.
\bibitem{Fritschel:2003qw}
  P.~Fritschel,
  arXiv:gr-qc/0308090.
  
 
\bibitem{Kuroda:2002bg}
  K.~Kuroda {\it et al.},
  Class.\ Quant.\ Grav.\  {\bf 19}, 1237 (2002).
 
 \bibitem{LISA}
 LISA Pre-Phase A Report, 2nd Ed. (1998):
http://www.srl.caltech.edu/lisa/documents/PrePhaseA.pdf
 
\bibitem{Crowder:2005nr}
  J.~Crowder and N.~J.~Cornish,
  Phys.\ Rev.\  D {\bf 72}, 083005 (2005).
  
\bibitem{Seto:2001qf}
  N.~Seto, S.~Kawamura and T.~Nakamura,
  Phys.\ Rev.\ Lett.\  {\bf 87}, 221103 (2001).


\bibitem{Asaka:1999xd}
  T.~Asaka and M.~Kawasaki,
  Phys.\ Rev.\  D {\bf 60}, 123509 (1999).
  
\bibitem{Giudice:2000ex}
  G.~F.~Giudice, E.~W.~Kolb and A.~Riotto,
  Phys.\ Rev.\  D {\bf 64}, 023508 (2001).
  
  
\bibitem{Kovner:1997ca}
  A.~Kovner, M.~A.~Shifman and A.~Smilga,
  Phys.\ Rev.\  D {\bf 56}, 7978 (1997).



\bibitem{Press:1989yh}
  W.~H.~Press, B.~S.~Ryden and D.~N.~Spergel,
Ap. J. 347, 590 (1989).
  
\bibitem{Coulson:1995nv}
  D.~Coulson, Z.~Lalak and B.~A.~Ovrut,
  Phys.\ Rev.\  D {\bf 53}, 4237 (1996).
 
\bibitem{Larsson:1996sp}
  S.~E.~Larsson, S.~Sarkar and P.~L.~White,
  Phys.\ Rev.\  D {\bf 55}, 5129 (1997).
  
  
\bibitem{Hindmarsh:1996xv}
  M.~Hindmarsh,
  Phys.\ Rev.\ Lett.\  {\bf 77}, 4495 (1996).
 
     \bibitem{AMSB} 
  L.~Randall and R.~Sundrum,
  Nucl.\ Phys.\ B {\bf 557}, 79 (1999);\\
  G.~F.~Giudice, M.~A.~Luty, H.~Murayama and R.~Rattazzi,
  JHEP {\bf 9812}, 027 (1998);\\
  J.~A.~Bagger, T.~Moroi and E.~Poppitz,
  JHEP {\bf 0004}, 009 (2000).
 
  

\bibitem{Abramovici:1992ah}
  A.~Abramovici {\it et al.},
  Science {\bf 256} (1992) 325.

\bibitem{recent-LIGO}
Abbott, B. et al. (The LIGO Scientific Collaboration) 
  Astrophys.\ J.\  {\bf 659}, 918 (2007).
  
 
\bibitem{Kawasaki:2004yh}
M.~Kawasaki, K.~Kohri and T.~Moroi,
Phys.\ Lett.\ B {\bf 625}, 7 (2005);
Phys.\ Rev.\ D {\bf 71}, 083502 (2005).

\bibitem{Kohri:2005wn}
  K.~Kohri, T.~Moroi and A.~Yotsuyanagi,
  %
  Phys.\ Rev.\ D {\bf 73}, 123511 (2006).
  
\bibitem{Jedamzik:2006xz}
  K.~Jedamzik,
  Phys.\ Rev.\  D {\bf 74}, 103509 (2006).
 
\bibitem{Moroi:1993mb}
  T.~Moroi, H.~Murayama and M.~Yamaguchi,
  Phys.\ Lett.\ B {\bf 303}, 289 (1993).
 
\bibitem{Bolz:1998ek}
  M.~Bolz, W.~Buchmuller and M.~Plumacher,
  Phys.\ Lett.\  B {\bf 443}, 209 (1998).
  
\bibitem{Bolz:2000fu}
    M.~Bolz, A.~Brandenburg and W.~Buchmuller,
    Nucl.\ Phys.\ B {\bf 606}, 518 (2001).
    
\bibitem{Ellis:2003dn}
  J.~R.~Ellis, K.~A.~Olive, Y.~Santoso and V.~C.~Spanos,
  Phys.\ Lett.\  B {\bf 588}, 7 (2004);
  L.~Roszkowski, R.~Ruiz de Austri and K.~Y.~Choi,
  JHEP {\bf 0508}, 080 (2005);
  D.~G.~Cerdeno, K.~Y.~Choi, K.~Jedamzik, L.~Roszkowski and R.~Ruiz de Austri,
  JCAP {\bf 0606}, 005 (2006).

\bibitem{Steffen:2006hw}
  F.~D.~Steffen,
  JCAP {\bf 0609}, 001 (2006).
 
\bibitem{Pradler:2006qh}
  J.~Pradler and F.~D.~Steffen,
  Phys.\ Rev.\  D {\bf 75}, 023509 (2007);
  Phys.\ Lett.\  B {\bf 648}, 224 (2007).


  
  
\bibitem{Spergel:2006hy}
  D.~N.~Spergel {\it et al.}  [WMAP Collaboration],
  Astrophys.\ J.\ Suppl.\  {\bf 170}, 377 (2007).
  
\bibitem{Mangano:2006ur}
  G.~Mangano, A.~Melchiorri, O.~Mena, G.~Miele and A.~Slosar,
  JCAP {\bf 0703}, 006 (2007).
  
  
\end{thebibliography}
\end{document}